\documentclass[aps,
prl,
reprint,
showkeys,
groupedaddress]{revtex4-2}
\usepackage{graphicx}% Include figure files
\usepackage{color}
\usepackage{xcolor}
\usepackage{amssymb}
\usepackage{setspace}
\usepackage[normalem]{ulem}
\usepackage{tabularx}
\usepackage{hyperref}% add hypertext capabilities

\begin{document}

\title{A Kinetic Criterion for Stokes-Einstein Relation Breakdown Based on Effective Collisional Geometry}

\author{Zhen-Wei Wu}%武振伟
\email{zwwu@bnu.edu.cn}
\affiliation{Institute of Nonequilibrium Systems, School of Systems Science, Beijing Normal University, 100875 Beijing, China}

\date{\today}

\begin{abstract}
Here we propose a kinetic framework for interpreting the Stokes-Einstein (SE) relation breakdown in supercooled liquids by introducing an effective collision diameter, $d_{\mathrm{eff}}$, derived from transport data. Numerical simulation of a model CuZr alloy reveal that $d_{\mathrm{eff}}$ increases upon cooling but saturates near the first peak of the radial distribution function just before SE breakdown. This saturation defines a geometric upper bound for the collisional cross-section beyond which further slowdown is governed by cooperative, heterogeneous motion rather than local collisional transport. Our analysis yields a compact criterion for SE breakdown in a mean-field perspective and provides physically interpretable inputs for future data-driven models of glassy dynamics.
\end{abstract}

\keywords{metallic glass-forming liquids; transport decoupling; kinetic-geometric criterion\\
PACS: 64.70.pe, 67.10.Jn, 05.20.Dd}

\maketitle

\noindent
{\it Introduction---}
The Stokes–Einstein (SE) relation is a classical result relating a particle's diffusion coefficient $D$ to the viscosity of the medium $\eta$ and the particle's hydrodynamic radius $R$ (often denoted $R_H$). In its simplest form for a spherical particle, it reads: $D=\frac{k_BT}{6 \pi \eta R_H}$, implying $D \eta / T$ is constant for a given particle size~\cite{binder2011glassy,brillo2011relation}. This relation works well for normal liquids at high temperatures and for mesoscopic tracer particles in a continuum solvent. However, as liquids are supercooled (cooled below their equilibrium melting point without crystallizing) and approach the glass transition~\cite{zhang2025mean,zhou2024k}, sizeable deviations from the SE relation appear~\cite{tarjus1995breakdown,lad2012signatures}. In many glass-forming liquids, the diffusion coefficient does not decrease as rapidly as the viscosity increases, leading to a ``decoupling'' of translational diffusion from viscosity and from structural relaxation times. In practical terms, molecules move faster (diffuse more easily) than one would predict from the vastly increased viscosity. This breakdown of the SE relation, often called a violation of Stokes–Einstein behavior, is now recognized as a hallmark of glassy dynamics in supercooled liquids~\cite{Han2011transport,Xu2009Appearance,wu2020revisiting,ren2025stokes}.

From a theoretical perspective, the breakdown of the SE relation in supercooled liquids signals the failure of the hydrodynamic assumptions at molecular length-scales. The SE relation was derived for a macroscopic or mesoscopic particle diffusing in a continuum fluid~\cite{brillo2011relation}. When the ``particle'' is a molecule in its own liquid, and especially as the liquid becomes highly viscous and structured, the assumptions underlying SE (e.g. a fixed effective radius and a homogeneous continuum drag) no longer hold. One way to describe the deviation is to say that the effective hydrodynamic radius of the diffusing molecules is not constant: if one forced the SE equation to hold, one would infer a hydrodynamic radius that grows or changes with cooling. In reality, this is a symptom of the breakdown of classical transport theory on the approach to the glass transition. I.e., in classical liquids the SE relation rationalizes diffusion with a single hydrodynamic radius, however, in supercooled liquids this mapping fails.

To bridge the microscopic and continuum descriptions of transport in glass-forming liquids, we introduce a transport-based effective collision diameter, $d_{\mathrm{eff}}(T)$, operationally defined by equating measured diffusion and relaxation times with their hard-sphere analogs. This construct enables a physically interpretable framework for quantifying how structure governs dynamics in supercooled liquids. By comparing $d_{\mathrm{eff}}$ against the nearest-neighbor spacing extracted from the pair correlation function, we uncover a distinct crossover marking the onset of SE breakdown. The resulting structural criterion not only provides a geometric ceiling for hydrodynamic transport mappings but also clarifies the emergence of dynamic heterogeneity. Through large-scale molecular dynamics simulations of a metallic glass-forming alloy, we demonstrate that this approach captures key signatures of the glass transition and yields insight into how local packing constraints govern macroscopic transport.

\vspace{5mm}

\noindent
{\it Result---}
We recognize that realistic glass-forming liquids are complex many-body systems in which particle interactions extend well beyond simple binary collisions. Nonetheless, it remains conceptually useful to approximate these interactions using a hard-sphere picture in order to elucidate the dominant transport mechanisms. Following Turnbull's classical argument for hard-sphere dynamics~\cite{turnbull1970free,turnbull1961free}, the self-diffusion coefficient $D$ can be expressed as
\begin{equation}
    D=\frac{1}{3} \bar{u} \lambda \bar{f}~,
\end{equation}
where $\bar{u}$ is the average particle speed, $\lambda$ is the mean free path, and $\bar{f}$ is a correlation factor that accounts for memory effects and many-body constraints ($\bar{f} \to 1$ in dilute fluids, and $\bar{f} \to 0$ in a perfect crystal or ideal glass). In deriving the effective collisional description, we adopt standard kinetic-theory expressions for dilute hard-sphere gases: the mean-free path is given by $\lambda = 1/(\sqrt{2}\,\pi d^2 n)$, and the binary collision rate per particle is
\begin{equation}
    \bar{Z} = \bar{u}/\lambda = \sqrt{2}\,\pi d^2 n \bar{u}~,
\end{equation}
where $d$ is the hard-core diameter and $n$ is the number density. The $\sqrt{2}$ factor originates from the relative velocity distribution and symmetry between identical particle pairs (see, e.g., Ref.~\cite{hansen2013theory,reif2009fundamentals}). To count effective encounters within a narrow tube around the instantaneous direction of motion (consistent with the elementary notion of a ``collision cylinder''), we use the one–component root-mean-square displacement over $\tau_\alpha$,
\begin{equation}
  \ell \equiv \ell_{\parallel} = \sqrt{\langle \Delta x^2 \rangle} = \sqrt{2D \tau_\alpha}~.
  \label{eq:ell_def}
\end{equation}
The corresponding number of traversed mean-free-path segments is then
\begin{equation}
    N_{\lambda} \equiv \frac{\ell}{\lambda} = \frac{\sqrt{2 D \tau_{\alpha}}}{\lambda}~.
\end{equation}
Substituting the expression for $\lambda$ from Eq.~(1), $\lambda = 3D / (\bar{u}\bar{f})$, yields
\begin{equation}
    N_{\lambda} = \frac{\bar{u}\bar{f}}{3D} \sqrt{2 D \tau_{\alpha}}~.
\end{equation}
An alternative estimate of the number of collisions over $\tau_\alpha$ is $N_Z = \bar{Z}\tau_\alpha$. Equating the two expressions and using Eq.~(2), we obtain
\begin{equation}
    \frac{\bar{u}\bar{f}}{3D} \sqrt{2 D \tau_{\alpha}} = \sqrt{2} \pi d^2 n \bar{u} \tau_{\alpha}~.
\end{equation}
Eliminating $\bar{u}$ and simplifying leads to
\begin{equation}
    \frac{d^2}{\bar{f}} = \frac{1}{3 \pi n} \frac{1}{\sqrt{D \tau_{\alpha}}}~.
\end{equation}
We define the transport-based effective collision diameter via $d_{\mathrm{eff}}^2 \equiv d^2/\bar{f}$. Using the per-particle volume $V_a \equiv 1/n$, this yields the final working expression:
\begin{equation}
    d^2_{\rm eff} = \frac{1}{3 \pi} \frac{V_a}{\sqrt{D \tau_{\alpha}}}~,
    \label{eq:work}
\end{equation}
which equates the observed transport properties to an effective hard-sphere picture, absorbing many-body corrections into a single temperature-dependent length scale.

For the numerical part we perform molecular dynamics (MD) simulations of a model Cu$_{50}$Zr$_{50}$ metallic glass-forming liquid using the LAMMPS package~\cite{Plimpton1995Fast}. Interatomic interactions are described by a realistic embedded-atom method (EAM) potential~\cite{mendelev2007using}, which accounts for many-body effects and has been widely validated for this alloy system~\cite{lad2017closely}. The simulated system consists of 10,000 atoms in a cubic box with periodic boundary conditions. Time integration is performed with a timestep of 2~fs. The system is first equilibrated at 2000~K under ambient pressure (NPT ensemble, $P=0$ bar) using a Nosé–Hoover thermostat and barostat for a duration of 4~ns. The liquid is then cooled to the target temperature at a rate of 1~K/ps and held for an additional 2~ns to ensure equilibration. Dynamical and structural properties are computed in the subsequent 2~ns production run. The self-diffusion coefficient $D$ is extracted from the long-time slope of the mean-squared displacement: $D = \lim \limits_{t \rightarrow \infty} \frac{1}{6t} \langle \Delta \mathbf{r}_j^2 (t) \rangle$, where $\Delta \mathbf{r}_j(t)$ denotes the displacement of particle $j$ over time $t$. The structural relaxation time $\tau_\alpha$ is determined from the self-intermediate scattering function~\cite{Kob1995Testing}, $F_{\rm s}(q,t) = N^{-1} \langle \sum_{j=1}^{N} \exp [{\rm i}\mathbf{q} \cdot \Delta \mathbf{r}_{j}(t)] \rangle$, as the time at which $F_{\mathrm{s}}(q,t)=e^{-1}$. The wavevector magnitude $q = 2.8~\text{\AA}^{-1}$ corresponds to the main peak of the static structure factor~\cite{Wu2018Stretched}, and the average is taken over all wavevectors with $|\mathbf{q}|=q$ and over initial configurations.

\begin{figure}[htb]
    \centering
    \includegraphics[width=0.95\linewidth]{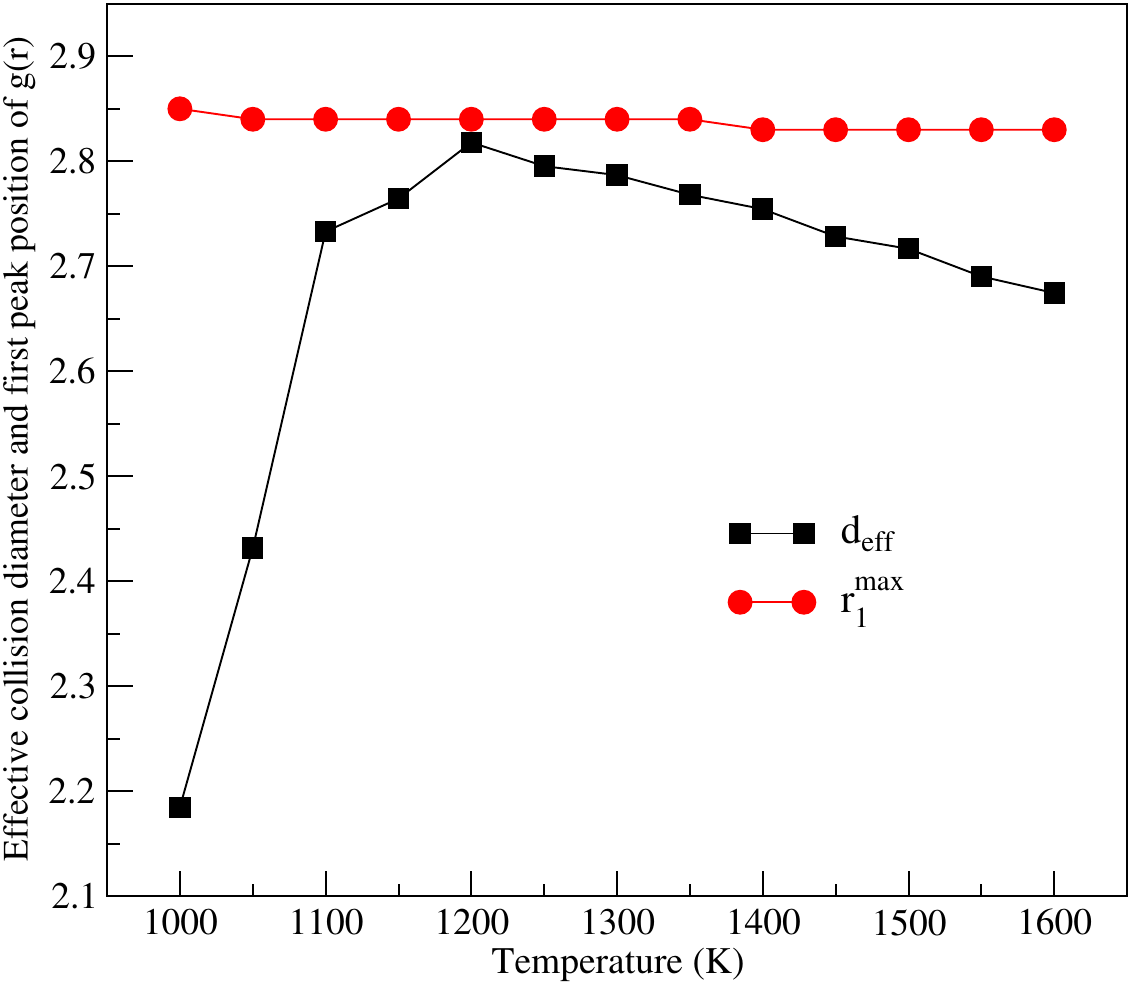}
    \caption{Effective collision diameter vs.~temperature. Squares: transport-inferred $d_{\mathrm{eff}}(T)$. Circles: nearest-neighbor spacing $r^{\mathrm{max}}_{1}(T)$ from global $g(r)$. Upon cooling, $d_{\mathrm{eff}}$ increases and approaches a ceiling set by $r^{\mathrm{max}}_{1}$. The approach of $r^{\mathrm{max}}_{1}(T)$ to this ceiling coincides with the onset of Stokes–Einstein behavior breakdown; further slowdown of $D$ and $\tau_\alpha$ is not captured by a single local length, and the inversion yields a turnover of $d_{\mathrm{eff}}$, reflecting the dominance of heterogeneous, activated dynamics.
    }
    \label{fig:d_effvsT}
\end{figure}

\begin{table*}[htb]
\caption{\label{tab:datall}Data used to compute the effective collision diameter $d_{\rm eff}$ and the scalar parameter $\mathcal{S}\ast$, based on diffusion, structural relaxation, per-particle volume, and pair correlation data at various temperatures.}
\begin{ruledtabular}
\begin{tabular}{ccccccc}
 & $D$ (\AA$^2$/ps) & $\tau_\alpha$(ps) & $V_a$ (\AA$^3$)
 & $d_{\rm eff}$ (\AA) & $r^{\rm max}_1$ (\AA) & $\mathcal{S}\ast$ \\
\hline
1600K &	0.27 & 0.29 & 18.89 & 2.67 & 2.83 & 1.12 \\
1550K &	0.24 & 0.32 & 18.84 & 2.69 & 2.83 & 1.11 \\
1500K &	0.21 & 0.35 & 18.79 & 2.72 & 2.83 & 1.09 \\
1450K &	0.18 & 0.40 & 18.75 & 2.73 & 2.83 & 1.08 \\
1400K &	0.15 & 0.46 & 18.70 & 2.75 & 2.83 & 1.06 \\
1350K &	0.12 & 0.55 & 18.65 & 2.77 & 2.84 & 1.05 \\
1300K &	9.80E-2 & 0.66 & 18.61 & 2.79 & 2.84 & 1.04 \\
1250K &	7.39E-2 & 0.86 & 18.56 & 2.80 &	2.84 & 1.03 \\
1200K &	5.37E-2	& 1.14 & 18.51 & 2.82 &	2.84 & 1.02\footnotemark[1] \\
1150K &	3.70E-2	& 1.78 & 18.46 & 2.76 &	2.84 & 1.06 \\
1100K &	2.28E-2	& 3.00 & 18.41 & 2.73 &	2.84 & 1.08 \\
1050K &	1.23E-2	& 8.87 & 18.36 & 2.43 &	2.84 & 1.36 \\
1000K &	5.37E-3	& 30.88 & 18.31 & 2.18 & 2.85 & 1.70 \\
\end{tabular}
\end{ruledtabular}
\footnotetext[1]{The onset of violation of SE behavior.}
\footnotetext[2]{All data given here are rounded to two decimal places.}
\end{table*}

Figure~\ref{fig:d_effvsT} presents the temperature dependence of the effective collision diameter $d_{\rm eff}$, as defined in Eq.~(\ref{eq:work}), for our three-dimensional metallic glass-forming system. At high temperatures, where the SE relation remains valid, $d_{\rm eff}$ decreases smoothly and nearly linearly with cooling. In this regime, the slope of the curve reflects the effective hydrodynamic radius of the constituent atoms. Upon further cooling, however, a clear turning point emerges, beyond which $d_{\rm eff}$ deviates from this trend, consistent with the breakdown of the SE relation and the emergence of heterogeneous, non-hydrodynamic transport.

Interestingly, Fig.~\ref{fig:d_effvsT} also reveals a gradually subtle increase in the first-neighbor peak position of the radial distribution function, $r_1^{\rm max}$, as temperature decreases (see also Table~\ref{tab:datall}). In metallic glass–forming systems, it is well documented that the first-neighbor peak position of the pair correlation function exhibits a modest increase with cooling (equivalently, a decrease with heating), reflecting subtle, composition-dependent reorganizations of short-range order in the liquid and glassy states. This trend has been reported in CuZr~\cite{ding2014temperature,dai2020x} and other metallic systems~\cite{tan2014correlation} via X-ray/neutron scattering~\cite{tan2014correlation,dai2020x} and MD simulations~\cite{ding2014temperature}, and discussed in recent overviews of metallic glass-formers~\cite{ding2017computational}. This ``anomalous'' expansion can be rationalized by considering the nature of the interatomic interactions in metallic liquids governed by many-body potentials. In systems described by EAM potentials, the effective strength of individual atomic ``bonds'' decreases as the local environment becomes increasingly crowded-an effect rooted in the Pauli exclusion principle. This behavior is fundamentally absent in purely pairwise potentials, which lack environment-dependent screening. As the supercooled liquid becomes more densely packed upon cooling, the increased coordination number weakens individual bonds and leads to a modest increase in the average bond length, thereby shifting $r_1^{\rm max}$ to larger values. The above argument is further supported by recent \emph{Ab initio} studies~\cite{jiang2025origin}.

As the liquid is cooled, a pronounced deviation from the high-temperature trend of $d_{\rm eff}(T)$ signals the onset of SE relation breakdown. While $d_{\rm eff}$ initially increases with decreasing temperature, reflecting the progressive tightening of dynamical cages, remarkably this growth is bounded above by the nearest-neighbor spacing from structure: $d_{\rm eff}(T) \lesssim\ r^{\rm max}_{1}(T)$ (defined by the first peak position of $g(r)$). Once $d_{\rm eff} \to r^{\rm max}_{1}$, the hydrodynamic mapping saturates: the system can no longer accommodate further increases in the collisional core without violating structural constraints. At this point, continued growth in the structural relaxation time $\tau_\alpha$ (or equivalently, the viscosity $\eta$) is no longer governed by binary, short-range collisional processes, but rather by collective, heterogeneous dynamics that emerge in the deeply supercooled regime.

Empirically, this saturation point coincides with the onset of SE breakdown. Beyond it, inversion of the transport expression can yield an apparent turnover in $d_{\rm eff}(T)$, where the inferred effective diameter decreases despite continued dynamical slowdown. This behavior reflects a breakdown in the assumption that long-time transport can be characterized by a single, local hydrodynamic length scale. Physically, the underlying picture is as follows: as cooling proceeds, the formation of ``tighter cages'' leads to more frequent short-range momentum transfers, which effectively mimic an increased collisional cross-section and thereby raise the inferred effective diameter, $d_{\rm eff}$. Importantly, the observed growth of $d_{\mathrm{eff}}$ with cooling in the pre-breakdown regime should not be interpreted as reflecting ``collective motion'' {\it per se}, but rather as an emergent signature of increasing short-range dynamical constraints. These constraints, arising from the gradual formation of cages, are naturally captured by our coarse-grained kinetic mapping. As the system approaches the onset temperature $T_{\mathrm{onset}}$, where the $F_{\rm s}(q,t)$ begins to exhibit a clear shoulder or incipient plateau following the ballistic regime, these short-range constraints become increasingly prominent. Even though the long-time dynamics remains diffusive, the frequent momentum exchanges at short times behave as if particles carry a larger collisional diameter. However, once these cages set the relevant contact scale, further slowdowns in $\tau_\alpha$ or viscosity $\eta$ arise from cooperative, heterogeneous rearrangements—not from further changes in local collisional geometry. In this regime, the notion of a single, well-defined hydrodynamic radius becomes ill-posed, and the extracted $d_{\mathrm{eff}}$ begins to decline accordingly.

Inverting our kinetic construction yields a compact and physically insightful criterion: $3\pi n d_{\mathrm{eff}}^{2}\sqrt{D\tau_{\alpha}} = 1$, which offers a valuable lens through which to interpret the breakdown of the SE relation. The left-hand side of this equation represents a dimensionless ``path-integrated cross section'' (or mean collision number) over a structural relaxation time $\tau_\alpha$: it combines the geometric cross-section density $n\pi d_{\mathrm{eff}}^{2}$, i.e., the expected number of collisional encounters per unit path length, with the diffusive displacement $\sqrt{D\tau_{\alpha}}$. Thus, the equality ``$=1$'' signifies that, during the time scale of $\tau_{\alpha}$, a typical particle undergoes approximately $\mathcal{O}(1)$ effective, uncorrelated collision, or equivalently, diffuses over a distance comparable to a single mean free path defined by $d_{\mathrm{eff}}$. The empirical observation that $d_{\rm eff}(T)$ approaches $r_1^{\max}(T)$ upon cooling motivates the introduction of a structural ceiling $d_\ast$ to organize the onset of SE breakdown. By replacing $d_{\mathrm{eff}}$ with the structural ceiling $d_{\ast} \simeq r^{\rm max}_1$ (a geometric upper bound for the collisional cross-section) in the kinetic criterion, we define a new scalar quantity
\begin{equation}
    \mathcal{S}\ast = 3\pi n\,(r^{\rm max}_1)^2 \sqrt{D \tau_\alpha}~.
    \label{eq:Sstar}
\end{equation}
As the liquid is cooled, $\mathcal{S}\ast$ approaches unity (see Table~\ref{tab:datall}). The condition $\mathcal{S}\ast \approx 1$ marks the onset of SE violation, where transport can no longer be rationalized by increasing a local collisional core because $d_{\mathrm{eff}}$ has saturated at the neighbor spacing. Beyond this point, continued increases of $\tau_\alpha$ accompanied by only modest changes in $D$ yield a turnover in $d_{\mathrm{eff}}$, signaling that long-time transport is now governed by collective dynamics rather than binary momentum exchange. This criterion also predicts clear trends: higher density (larger $n$) or larger probe size (larger $d_\ast$) shift the onset of SE violation to higher temperatures, whereas strong liquids, which exhibit weaker growth in $\sqrt{D\tau_\alpha}$, display a more delayed and gradual breakdown, consistent with empirical observations.

\begin{figure}[htb]
    \centering
    \includegraphics[width=\linewidth]{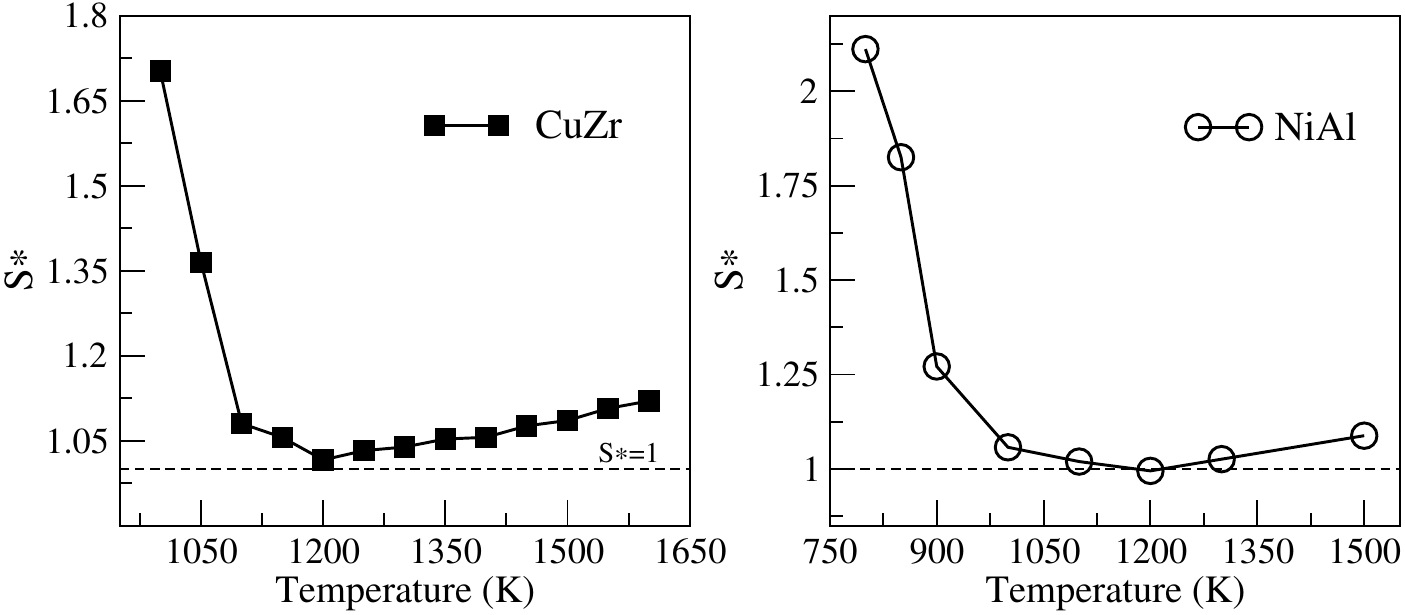}
    \caption{Temperature dependence of the dimensionless kinetic–geometric quantity $\mathcal{S}\ast=3\pi n\, (r_1^{\max})^2 \sqrt{D\tau_\alpha}$ for two metallic glass-forming alloys, CuZr and NiAl. For both systems, upon cooling from the high-temperature liquid, $\mathcal{S}\ast$ develops a minimum close to unity, $\mathcal{S}\ast \simeq 1$, at the temperature where the transport-inverted effective diameter $d_{\mathrm{eff}}$ saturates at the structural neighbor spacing.
    }
    \label{fig:star}
\end{figure}

To probe the robustness of the proposed criterion beyond a single composition, we have performed an additional set of simulations for a Ni$_{50}$Al$_{50}$ metallic alloy using the same protocol described above for Cu$_{50}$Zr$_{50}$. For this second system we determined the measurable inputs $\{D,\tau_\alpha,n,g(r)\}$ and computed the dimensionless quantity $\mathcal{S}\ast$. The resulting $\mathcal{S}\ast$ at temperatures considered for both alloys is shown in Fig.~\ref{fig:star}. Despite their different chemistry and interaction parameters, the two systems exhibit quite similar qualitative trends: upon cooling from the high-temperature liquid, $\mathcal{S}\ast$ approaches a minimum value close to unity, $\mathcal{S}\ast \simeq 1$, at the temperature where the SE relation starts to fail. This is precisely the regime in which the transport-inverted effective diameter $d_{\mathrm{eff}}$ saturates at the structural neighbor scale, so that further increases in $\tau_\alpha$ (or $\eta$) can no longer be attributed to a growing local collisional core. This cross-check on Ni$_{50}$Al$_{50}$ indicates that the kinetic–geometric criterion based on $\mathcal{S}\ast$ is not specific to Cu$_{50}$Zr$_{50}$, but captures a more general structural–dynamical correspondence in metallic glass-forming liquids. It reinforces the view that the condition $\mathcal{S}\ast \approx 1$ provides a concrete, experimentally accessible marker for the onset of SE breakdown, and thus a useful/promising target for future numerical and experimental tests across different alloy families. Within our kinetic-geometric criterion framework, the data shown in Fig.~\ref{fig:star} also admits a simple kinetic interpretation. The $\mathcal{S}\ast$ can be regarded as the effective number of \emph{uncorrelated} collisions available to a particle during one structural relaxation time $\tau_{\alpha}$. Well above the SE-breakdown temperature this number exceeds unity, so that several such uncorrelated collisional events can, in principle, occur within $\tau_{\alpha}$. Upon supercooling, however, the maximal number of effective uncorrelated collisions allowed on the $\tau_{\alpha}$ time scale decreases and approaches one as $\mathcal{S}\ast \to 1$. Beyond this point, no further gain in uncorrelated collisions can be achieved by tuning $d_{\mathrm{eff}}$ resulted from the geometric constraint; collisions occurring within $\tau_{\alpha}$ are then necessarily embedded in correlated, cooperative rearrangements rather than being independent binary events.

\vspace{5mm}

\noindent
{\it Discussion---}
In this work, we operationally define the effective collision diameter, $d_{\mathrm{eff}}(T)$, by equating the measured (or simulated) transport to its hard-sphere analog. Across all state points, $d_{\mathrm{eff}}(T)$ increases slightly at first upon cooling but yet remains bounded above by the average nearest-neighbor spacing $r^{\rm max}_1(T)$, extracted from the first peak of the radial distribution function $g(r)$. This ceiling behavior serves as a structural constraint that becomes particularly relevant in the supercooled regime.

From a phenomenological perspective, the saturation of $d_{\mathrm{eff}}$ provides a simple yet powerful explanation for the decoupling between translational diffusion and structural relaxation. As cooling proceeds, the diffusion coefficient $D$ decreases less rapidly than $\eta$ or $\tau_\alpha$. This is because particles no longer diffuse via independent, local collisions but instead through cooperative rearrangements involving correlated clusters of atoms. In this regime, some particles are transiently shielded from collisions by their neighbors, effectively reducing the number of collisional events detrimental to transport. Viewed through an effective-theory lens, this corresponds to a cessation in the growth of $d_{\mathrm{eff}}$, which in turn facilitates atomic diffusion and leads to a natural breakdown of the SE relation. The ceiling behavior of $d_{\mathrm{eff}}$ thus points to a key role of local structure in setting the conditions for dynamic heterogeneity and cooperative motion near the glass transition. It implies that transport anomalies such as decoupling and non-Arrhenius dynamics, are not merely kinetic in origin but are instead deeply tied to geometric and packing constraints within the liquid.

It is useful to place our kinetic construction in the context of existing diagnostics of Stokes–Einstein breakdown. In particular, the product $D\tau_\alpha$ (or, equivalently, $D\eta/T$ when viscosity data are available) has long been employed as a convenient indicator of SE violation in glass-forming liquids, and has been used, for example, to identify kink temperatures in metallic glass-formers~\cite{Han2011transport}. In this sense, our work does not introduce a completely new diagnostic variable but embeds $D\tau_\alpha$ into a more structured kinetic–geometric framework, i.e., provides an effective single-particle collisional picture of the physics underlying the SE relation and its breakdown. By combining Turnbull's hard-sphere–inspired kinetic picture with measured $(D,\tau_\alpha,n)$ and structural information from $g(r)$, we obtain: (i) a dimensionless collisional quantity (a ``mean collision number'' accumulated over $\tau_\alpha$); (ii) a structural indicator set by the nearest-neighbor length scale $r_1^{\max}$ extracted from $g(r)$; and (iii) a predictive onset marker $\mathcal{S}\ast \equiv 3\pi n (r_1^{\max})^2 \sqrt{D\tau_\alpha} \simeq 1$ that explains not only that diffusion–viscosity decoupling appears, but also why and when it appears, in terms of a saturation of the effective collisional cross-section at a structural ceiling. In this way, our criterion extends the traditional use of $D\tau_\alpha$ by explicitly tying it to the underlying structure and by clarifying its parameter dependencies (density, effective probe size, and fragility). Importantly, the proposed dimensionless quantity $\mathcal{S}\ast$ is constructed entirely from experimentally or numerically accessible observables $\{D,\tau_\alpha,n,g(r)\}$, and the prediction that $\mathcal{S}\ast \approx 1$ at the onset of SE violation provides a concrete target for future experimental tests.

Finally, our results also hold implications for emerging data-driven approaches. Recent studies have demonstrated that machine-learning (ML) models, when combined with physically motivated descriptors~\cite{boattini2020autonomously,boattini2021averaging}, can successfully capture glassy dynamics. The criterion introduced in Eq.~(\ref{eq:Sstar}) offers such a descriptor: a physically interpretable scalar that encodes collisional opacity over structural timescales. As $d_{\mathrm{eff}}$ connects structure to transport through a single, compact variable, it is well-suited for integration into ML pipelines aimed at predicting dynamics or identifying relevant collective degrees of freedom. Incorporating such physically meaningful priors could enhance both the accuracy and interpretability of data-driven models for disordered materials.\\

%\vspace{5mm}

\noindent
{\it Acknowledgments---}
Z.W.W. thanks the members of the ``Beijing Metallic Glass Club'' for the long-term fruitful discussions. This work was supported by the National Natural Science Foundation of China (Grant Nos.~12474184, 52031016, and 11804027).

\bibliography{ref}
\bibliographystyle{iopart-num.bst}

\end{document}